%% file: main.tex
\documentclass[12pt,number,sort&compress]{elsarticle}

\usepackage{lineno}

\journal{Nuclear Instrumentation and Methods in Physics Research A}

\newcommand{\gsim}{\hbox{ \raise3pt\hbox to 0pt{$>$}\raise-3pt\hbox{$\sim$} }}
\newcommand{\lsim}{\hbox{ \raise3pt\hbox to 0pt{$<$}\raise-3pt\hbox{$\sim$} }}
\newcommand{\del}{\ifmmode{\nabla}               \else{$\nabla$ }               \fi}

\newcommand{\figdir}{.}
\usepackage{graphicx}

\usepackage{amssymb}
\usepackage{amsmath}

\begin{document}

\begin{frontmatter}



\title{Cosmic ray tests of a GEM-based TPC prototype \\
                   operated in Ar-CF$_4$-isobutane gas mixtures: II}

%
%
%
\input{authorlist.tex}


%
%
%
%
\begin{abstract}
The spatial resolution along the pad-row direction was measured 
with a GEM-based TPC prototype for the future linear collider
experiment in order to understand its performance for tracks 
with finite projected angles with respect to the pad-row normal.
The degradation of the resolution due to the angular pad effect
was confirmed to be consistent with the prediction of a simple
calculation taking into account the cluster-size distribution
and the avalanche fluctuation. 
\end{abstract}
%
%
%
%
%

\end{frontmatter}


%
%
%
%
%
%

\section{Introduction}

In the previous paper \cite{ref1} we demonstrated the feasibility of a 
GEM-based Time Projection Chamber (TPC) operated in an
Ar-CF$_4$-isobutane gas mixture as a central tracker (LCTPC) for the future
linear collider experiments (ILC \cite{refLC1} and CLIC \cite{refLC2}).
The spatial resolution along the pad-row direction was presented 
for tracks nearly perpendicular to the pad row in Ref. \cite{ref1} 
because the resolution in the $r$-$\phi$ plane better than
$\sim$ 100 $\mu$m per pad row
for stiff and radial tracks is of prime importance for
the physics goals of the experiments.
A TPC equipped with GEM readout is certainly an ideal main tracker,
which is free from the $E \times B$ and the angular wire effects
inherent in conventional TPCs with MWPC readout.

It should be noted, however, that the azimuthal resolution degrades with
increasing projected track angle ($\phi$) measured from the pad-row normal
because of the angular pad effect
as far as conventional pads are employed for readout.
As will be seen the angular pad effect adds an almost constant offset to
the resolution with its amount depending on the pad height as well as
the track angle.
Therefore the requirement for the spatial resolution above would not be
met for inclined tracks.  
The degraded resolution for slanted and/or low-momentum tracks
provided by the central tracker could affect the physics capability
of the whole detector system.
An example is the need for good energy resolution for soft jets;
thus it is important to understand whether such things can be
affected by the design of the TPC.

In this paper the resolutions measured
with cosmic rays for inclined tracks 
in a prototype TPC are presented and compared to the expectation
in order to provide a basis for the optimization of the pad height of
the LCTPC.  

The expected deterioration of the resolution for inclined tracks,
compared to that for right angle tracks, is estimated in Section 2.
The comparison of the measured resolution with the expectation is
presented in Section 3 after a brief description of the experiment.
Section 4 is devoted to a discussion and 
Section 5 concludes the paper.

\section{Expectation}

For right angle tracks ($\phi = 0^\circ$) 
the resolution along the pad-row direction ($\sigma_{\rm X}$) is
approximately given by
\begin{linenomath}
\begin{equation}
\sigma_{\rm X}^2 = \sigma_{\rm X00}^2 + \frac{D^2}{n_{\rm eff}} \cdot z 
\end{equation}
\end{linenomath}
where
$\sigma_{\rm X00}$ is the intrinsic resolution\footnote{
The values of $\sigma_{\rm X00}$ are measured to be about 100 $\mu$m without axial
magnetic field ($B$ = 0 T) and $\sim$ 50 $\mu$m for $B$ = 1 T. 
The observed $B$-dependence of $\sigma_{\rm X00}$ is most likely due to 
the intrinsic track width.
See Appendix C of Ref.~\cite{ref1} for the possible contributors to
the intrinsic term.
},
$D$ is the diffusion constant, $n_{\rm eff}$ is the effective
number of electrons per pad row, and $z$ is the drift distance
\cite{ref2}\footnote{
The finite pad-pitch term \cite{ref1} is neglected here. 
}.
It is worth noting that the value of $n_{\rm eff}$ is almost independent
of the drift distance \cite{ref3}\footnote{
In Refs. \cite{ref1,ref2,ref3}, $n_{\rm eff}$ is denoted as $N_{\rm eff}$,
which is reserved for the effective number of {\it clusters} per pad row
(see below) in the present paper.
}.
Even in the case of finite track angle
the explicit drift-distance dependence (the second term) of the resolution is 
scarcely affected by practically small $\phi$ \cite{ref4}.
The first term is, on the other hand, sensitive to the track angle.
It may be expressed as
\begin{linenomath}
\begin{equation}
\sigma_{\rm X0}^2 = \sigma_{\rm X00}^2 + \frac{h^2 \cdot \tan^2 \phi}
{12 \cdot N_{\rm eff}} 
\end{equation}
\end{linenomath}
where $h$ is the pad height\footnote{
More precisely $h$ should be understood as the pad-row pitch,
which is usually slightly larger than the pad height
when the readout plane is covered over with pads.
The pad-row pitch and the pad height ($h$) are not distinguished in the 
present paper.
}
and $N_{\rm eff}$ is the effective number
of {\it clusters} per pad row (see footnote 3).
The second term in Eq. (2) represents the contribution of the
angular pad effect to the resolution,
which is parametrized by $N_{\rm eff}$.

In fact, $N_{\rm eff}$ is a function of $\phi$, $\theta$, $z$ and $h$:
\begin{linenomath}
\begin{equation}
N_{\rm eff} = N_{\rm eff}(\phi, \theta, z, h) 
\end{equation}
\end{linenomath}
where $\theta$ is the angle between the track and the readout pad
plane\footnote{
$\theta$ is defined to be 0$^\circ$ when the track is parallel to the
readout plane.
}.
Let us consider first the $h$ dependence of $N_{\rm eff}$,
i.e. $N_{\rm eff}(0, 0, z, h)$. 
The average number of clusters per pad row ($\left< N \right>$) is
proportional to $h$.
Furthermore the $z$ dependence of the effective
number of clusters
due to de-clustering
is expected to be small \cite{ref4}\footnote{
The value of $\sigma_{\rm X0}$ given by Eq. (2) is therefore
practically independent of $z$. 
}.
Accordingly
\begin{linenomath}
\begin{equation}
  N_{\rm eff}(0, 0, z, h) \sim N_{\rm eff}(0, 0, 0, h)
                         = N_{\rm eff}(\left< N \right>) \;.
\end{equation}
\end{linenomath}

For a fixed number of clusters $N$, $N_{\rm eff}$ is given by
\begin{linenomath}
\begin{equation}
N_{\rm eff}(N) = \left< \sum_{i=1}^N Q_i^2 \; / 
                     \left( \sum_{i=1}^N Q_i \right) ^2 \right> ^{-1} 
\end{equation}
\end{linenomath}
where $Q_i$ is the total charge of the cluster $i$ given by
\begin{linenomath}
\begin{equation}
Q_i = \sum_{j=1}^{n_i} q_j 
\end{equation}
\end{linenomath}
with $q_j$ being the amplified signal of the $j$-th electron in the
cluster $i$ of size $n_i$ (see Appendix).
$N_{\rm eff}$ was estimated by numerical calculations, taking into account 
the cluster-size distribution for argon \cite{ref5}, and is shown 
in Fig. 1, with (filled circles) and without (open circles) the typical
avalanche fluctuation (a Polya distribution with $\theta =
0.5$\footnote{
The parameter $\theta$ for Polya distributions
(see, for example, Ref. \cite{ref2})
should not be confused with
the track angle $\theta$ defined above.
We use the same symbol since they can be easily distinguished by
their units.
})
for each electron.
The figure tells us that the effective number of clusters
is considerably smaller than $N$ because of the large cluster-size fluctuation.
Furthermore it is not a linear function of $N$; 
see Appendix for a qualitative estimation of $N_{\rm eff}$.
In real case, $N$ is not a constant and obeys Poisson statistics
with the average $\left< N \right>$. The curves in Fig. 1 show the
effective number of clusters as a function of $\left< N \right>$.
\begin{figure}[htbp]
\begin{center}
\includegraphics*[scale=0.60]{\figdir/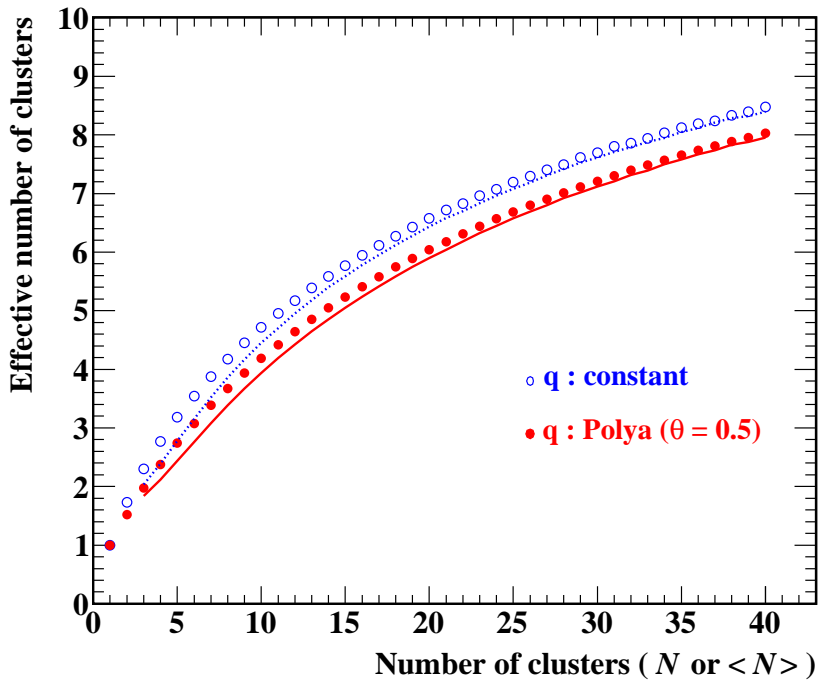}
\end{center}
\vspace{-10mm}
\caption{\label{fig1}
\footnotesize Effective number of clusters ($N_{\rm eff}$) as a function of the
total or average number of clusters: plots for fixed $N$, and curves for
Poissonly distributed $N$. The filled (open) circles and the full (dotted)
curve are calculated with (without) the avalanche fluctuation.}
\end{figure}

From the curve with the avalanche fluctuation in Fig. 1
the effective number of clusters for a given track angle can be
estimated since
\begin{linenomath}
\begin{equation}
 \left< N \right> = \frac{d \cdot h}{\cos \phi \cdot \cos \theta} 
\end{equation}
\end{linenomath}
with $d$ being the cluster density
($\sim$ 2.43/mm for minimum ionizing particles in argon \cite{ref6}),
and
\begin{linenomath}
\begin{equation}
N_{\rm eff}(\phi, \theta, 0, h) 
= N_{\rm eff}(\left< N \right>) 
= N_{\rm eff}\left( \frac{d \cdot h}{\cos \phi \cdot \cos \theta} \right) \;.
\end{equation}
\end{linenomath}
Let us define $S_{\rm X00}$ as the square root of the second term in
Eq. (2) at $z$ = 0:
\begin{linenomath}
\begin{equation}
 S_{\rm X00} \equiv \frac{h \cdot \tan \phi}
             {\sqrt{12 \cdot N_{\rm eff}(\left< N \right>)}}\;.
\end{equation}
\end{linenomath}
Fig. 2 shows   $S_{\rm X00}$  as a function of the pad height ($h$)
for $\phi$ = 5$^\circ$, 10$^\circ$, 15$^\circ$ and 30$^\circ$,
calculated with $\theta$  fixed to 0$^\circ$.
It should be noted that the resolutions shown in the figure are
the best possible values expected to be obtained without diffusion (at
$z$ = 0) and without contribution of the intrinsic term
($\sigma_{\rm X00}$).

\begin{figure}[htbp]
\begin{center}
\includegraphics*[scale=0.54]{\figdir/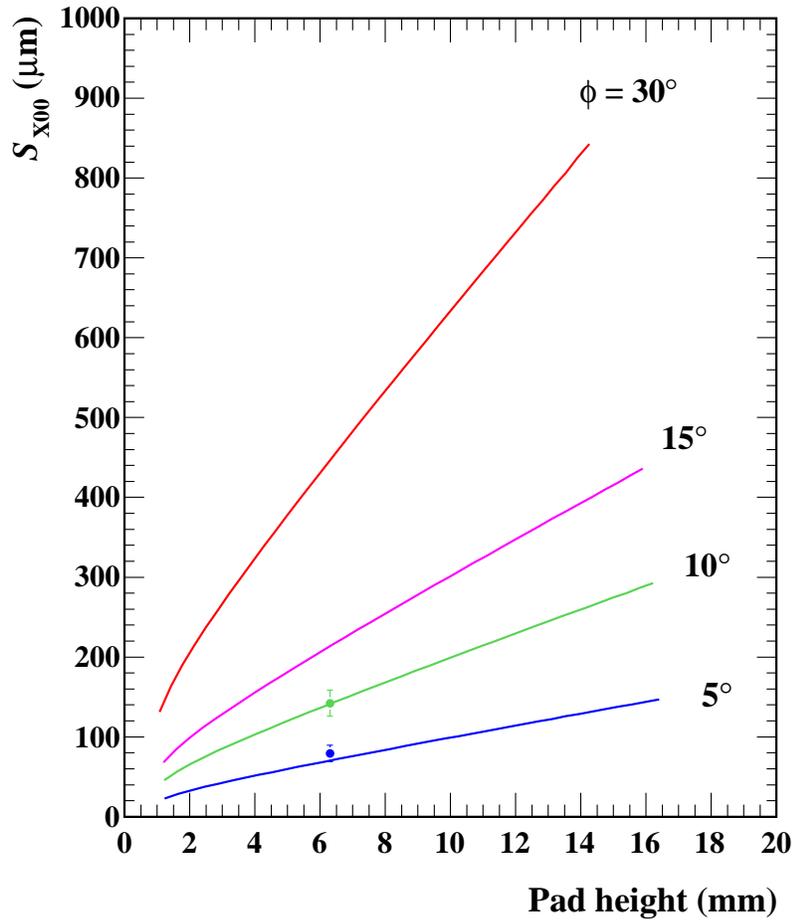}
\end{center}
\vspace{-10mm}
\caption{\label{fig2}
\footnotesize Expected contribution of the angular pad effect
 ($S_{\rm X00}$) as a function of the pad height ($h$) for different track angles.
Poisson statistics is assumed for the number of clusters and
a Polya distribution ($\theta = 0.5$) is assumed for the avalanche fluctuation.
The tracks are assumed to be minimum ionizing and parallel to the readout plane.
The points plotted at $h$ = 6.3 mm are the measurements (see Section 3.2).}
\end{figure}

\section{Experiment}
 \subsection{Setup and analysis}

We used a small GEM-based TPC prototype (MP-TPC) operated in
a gas mixture of Ar (95\%)-CF$_4$ (3\%)-isobutane (2\%)
at atmospheric pressure.
The MP-TPC is a small time projection chamber with a maximum
drift length of 257 mm.
Its gas amplification device is a triple GEM, 100 mm $\times$ 100 mm
in size. 
The amplified electrons are collected by a readout plane 
placed right behind the GEM stack, 
having 16 pad rows at a pitch ($h$) of 6.3 mm, each consisting of
1.17 mm $\times$ 6 mm rectangular pads arranged at a pitch of 1.27 mm.
The neighboring pad rows are staggered by half a pad pitch.
The pad signals are then fed to readout electronics, a combination of
preamplifiers, shaper amplifiers and digitizers.  
See Ref. \cite{ref1} for details of the experimental setup and the analysis
procedure for the cosmic ray tests of the MP-TPC.

We re-analyzed the data taken for the previous paper on the normal
incident tracks \cite{ref1} with different cuts on the track angles.
Among the data sets the data collected with a drift field of 250 V/cm
and $B$ = 0 T were selected because of its highest statistics and the negligible
influence of the finite pad-pitch term
in the absence of axial magnetic field \cite{ref1}.
The offset to the resolution due to finite
track angle is to be added quadratically as well in the presence of
a magnetic field, depending on the local track angle, 
at drift distances where the finite pad-pitch term is negligible.

The track angle distributions are shown in Fig. 3.
As mentioned in Introduction our primary concern in the cosmic ray tests with
the MP-TPC was the resolution for right angle tracks.
Therefore the acceptance to inclined tracks was limited by 
trigger-counter arrangement in order to reduce the trigger rate to
the relatively slow readout electronics.
The maximum available track angle is thus $|\phi| \lsim 10^\circ$
as seen in Fig. 3 (a).
\begin{figure}[htbp]
\begin{center}
\includegraphics*[scale=0.80]{\figdir/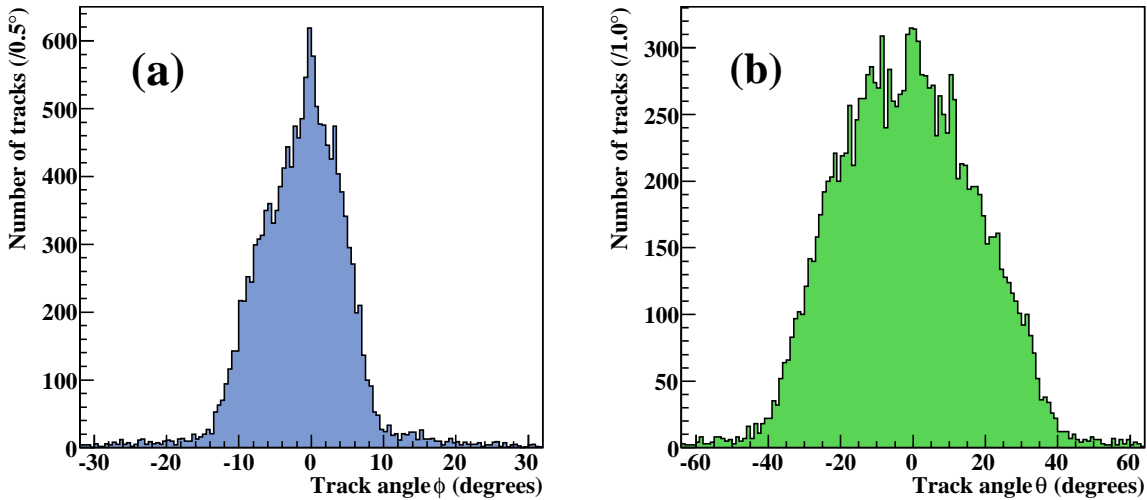}
\end{center}
\vspace{-10mm}
\caption{\label{fig3}
\footnotesize Track angle distributions: (a) for $\phi$, and (b) for $\theta$.}
\end{figure}

 \subsection{Results}

The spatial resolutions along the pad-row direction are shown in 
Fig. 4 for $|\phi|$ = 0$^\circ$, 5$^\circ$ and 10$^\circ$,
for tracks nearly parallel ($|\theta| \leqq 10^\circ$) to the readout plane.
The azimuthal angle cuts are the nominal values $\pm$ 2$^\circ$.
The resolutions squared as function of the drift distance ($z$) were fitted by a
function
$\sigma_{\rm X}^2 = \sigma_{\rm X0}^2 + D^2 / n_{\rm eff} \cdot z$
for free parameters $\sigma_{\rm X0}$ and $n_{\rm eff}$,
with the value of $D$ fixed to 315 $\mu$m/$\sqrt{\rm cm}$
given by Magboltz \cite{ref7}.
$S_{\rm X00}$ and $N_{\rm eff}$ were then obtained using Eqs. (2) and
(9) for each $\phi$,
assuming $\sigma_{\rm X00}$ to be  $\sigma_{\rm X0}$ measured
for $\phi$ = 0$^\circ$.
The resultant $\sigma_{\rm X0}$, $n_{\rm eff}$, $S_{\rm X00}$ and $N_{\rm eff}$ 
are summarized in Table~1 along with  the values of
$S_{\rm X00}$ and $N_{\rm eff}$ calculated for $h$ = 6.3 mm.
The measured values of $S_{\rm X00}$ are plotted also in Fig. 2.

The measured values of $S_{\rm X00}$ and
$N_{\rm eff}$ are consistent with those given by the calculation.
In addition,
the values of $n_{\rm eff}$ for inclined tracks are close to that for
normal incident tracks as expected and are consistent with
an estimation in Ref. \cite{ref2}.

\begin{figure}[htbp]
\begin{center}
\includegraphics*[scale=0.86]{\figdir/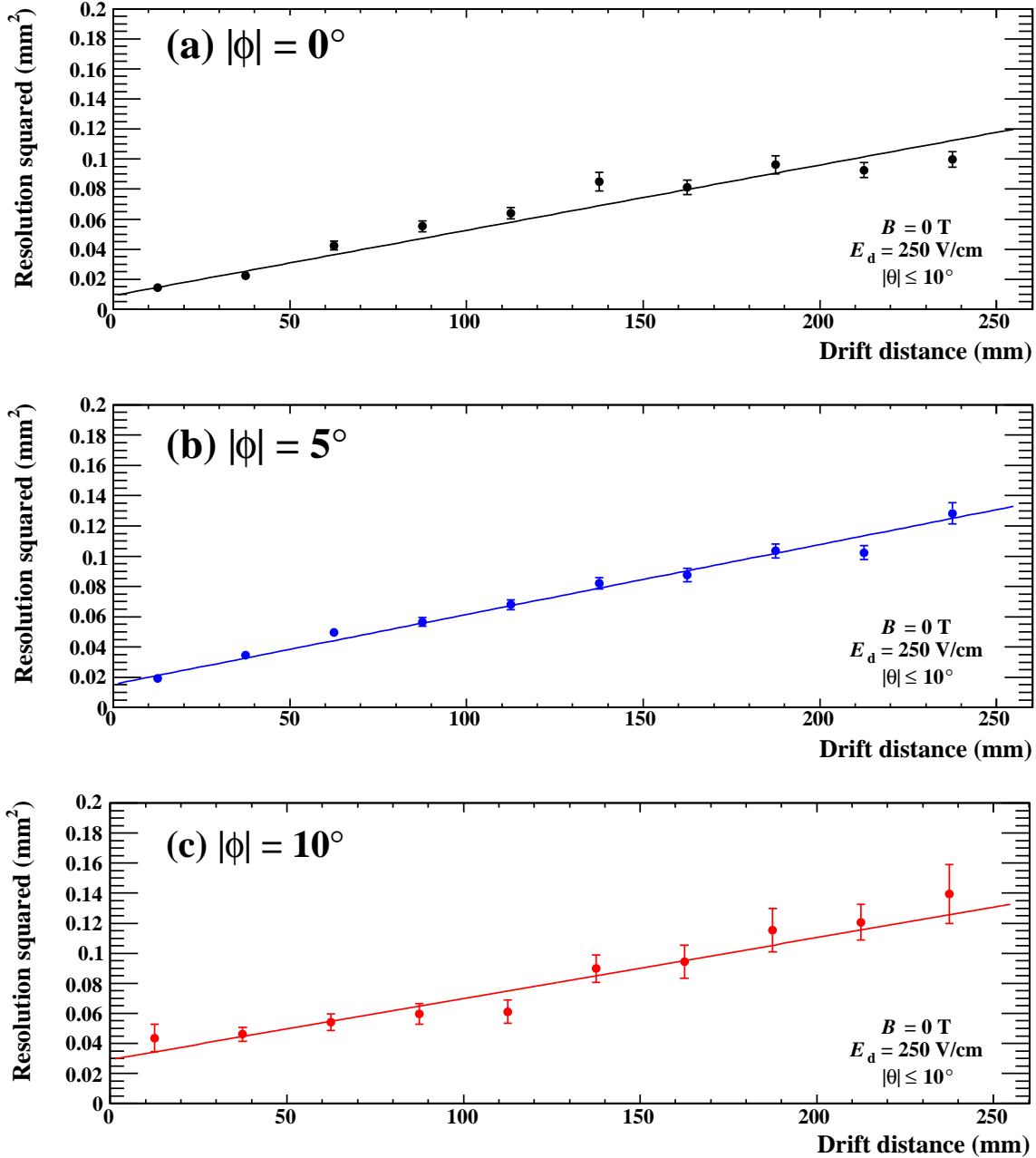}
\end{center}
\vspace{-10mm}
\caption{\label{fig4}
\footnotesize Resolution squared ($\sigma_{\rm X}^2$) as a function
of the drift distance ($z$):
(a) for $|\phi|$ = 0$^\circ$, (b) for $|\phi|$ = 5$^\circ$ and
(c) for $|\phi|$ = 10$^\circ$. See text for the straight lines 
fitted through the data points.}
\end{figure}

\input{Table1.tex}

\section{Discussion}

The figure of merit for the azimuthal spatial resolution of a
cylindrical TPC is the resolution per projected track length
in the $r$-$\phi$ plane along the radial direction.
From Eqs. (1), (2) and (9), the resolution per pad row is expressed as
\begin{linenomath}
\begin{equation}
\sigma_{\rm X}^2 \sim \sigma_{\rm X00}^2 + S_{\rm X00}^2
                         + \frac{D^2}{n_{\rm eff}} \cdot z 
\end{equation}
\end{linenomath}
at drift distances where the finite pad-pitch term is negligible,
and each of the three terms is a function of the pad height $h$.

We consider here the effect of splitting a pad row with $h$ = $H$
into a couple of identical pad rows with a height of $H$/2.
Let us assume for simplicity that the combined track coordinate is 
given by the average of the two measurements
provided by the neighboring pad rows with $h$ = $H$/2.
Then the resolution per projected track length $H$ becomes
\begin{linenomath}
\begin{equation}
\sigma_{\rm X}^2 = \frac{{\sigma_{\rm X}^\ast}^2}{2}
\end{equation}
\end{linenomath}
where $\sigma_{\rm X}^\ast$ is the resolution obtained with a 
single pad row with the halved height.
The diffusion contribution (the third term in Eq.(10)) is almost
unaffected since $n_{\rm eff}$ is approximately
proportional to the pad height \cite{ref3}.

On the other hand, the angular pad effect ($S_{\rm X00}$) is 
reduced appreciably.
We temporarily assume $N_{\rm eff}(\left< N \right>)$ to be proportional to
the pad height\footnote{
This is a bolder assumption than $n_{\rm eff} \propto h$ above
(see Fig. 1).
}.
Then
\begin{linenomath}
\begin{equation}
 S_{\rm X00}^2 \propto h 
\end{equation}
\end{linenomath}
from Eq. (9), and the combined contribution of the angular pad effect
($S_{\rm X00}$) per projected track length $H$ is halved
because of Eq. (11).
Actually $S_{\rm X00}$ is reduced by more than a factor of 2 because
Eq. (12) gives an overestimate for a smaller $h$ (see Fig. 2).

In addition, $\sigma_{\rm X00}^2$ at long drift distances can be shown
mathematically to be 
\begin{linenomath}
\begin{equation}
\sigma_{\rm X00}^2 = \frac{B_0^2}{n_{\rm eff}}
\end{equation}
\end{linenomath}
with a constant $B_0$ independent of the pad height
if the contribution of the electronic noise is negligible
(see Appendix B of Ref. \cite{ref1}).
Similarly to the diffusion contribution, the intrinsic term
($\sigma_{\rm X00}$) in the combined resolution is expected to be
close to the counterpart in the resolution for a single pad row with $h$ = $H$. 

Consequently the net effect of halving the pad height on the resolution
per projected track length is essentially the alleviation of the angular pad effect
($S_{\rm X00}$) by more than a factor of 2.
For example, Eq. (9) gives $S_{\rm X00}$ $\sim$ 140 $\mu$m (450 $\mu$m)
for $\phi$ = 10$^\circ$ (30$^\circ$) with $h$ = 6.3 mm,
while the corresponding value for a couple of pad rows 
with $h$ = 3.15 mm is about 60 $\mu$m (200 $\mu$m).
The spatial resolutions, and therefore the momentum resolutions
improve significantly for slanted and/or low momentum tracks
with the shorter pads.  

The number of voxels in the sensitive volume of a TPC is doubled
if the pad height is halved (with the electronics channel density doubled). 
This would enhance the pattern recognition capability and
the d$E$/dx resolution of the TPC as well.

\section{Conclusion}

The azimuthal spatial resolutions for inclined tracks were measured
with a GEM-equipped prototype TPC as well as for right angle tracks.
The angular pad effect contributes as a virtually constant offset
to the spatial resolution to be added quadratically,
depending on the track angle and the pad height.
The offsets are found to be consistent with the predictions given by
a simple model calculation taking into account the cluster-size
distribution and the avalanche fluctuation.

The results are expected to be useful in optimizing the pad height of the
LCTPC from the physics point of view.

\section*{\nonumber Acknowledgments}
We would like to thank the group at the KEK cryogenics science center for
the preparation and the operation of the superconducting magnet.
We are also grateful to many colleagues of the LCTPC collaboration for their
continuous encouragement and support, and for fruitful discussions.
This work was supported by the Creative Scientific Research
Grant No. 18GS0202 and No. 23000002
of the Japan Society of Promotion of Science.

\appendix

\section{Behavior of $N_{\rm eff}(N)$}

The effective number of clusters ($N_{\rm eff}$) parametrizes
the degradation of the resolution due to the angular pad effect
(the second term in Eq. (2)).
We consider here the behavior of $N_{\rm eff}$ at $z = 0$ 
as a function of the fixed total number of clusters per pad row ($N$),
qualitatively for $N$ = 1, 2, 4 and $\infty$.
The track coordinate ($X$) along the pad-row direction is assumed 
to be determined from the charge centroid 
of the clusters detected by the pad row having an infinitesimal
pad pitch.
The clusters are fully intact at $z = 0$
and are assumed to be point-like.

In order to estimate $N_{\rm eff}(N)$
it is necessary to evaluate the variance ($\equiv {S_N}^2$) 
of the charge centroid of
$N$ clusters, each with charge $Q_i$ and coordinate $x_i$,
which are randomly scattered over the lateral range on the pad row
covered by an inclined track ($h \cdot \tan \phi$).

\begin{enumerate}
\item $N = 1$ \\
      The resolution ($\equiv S_1$) does not depend on the cluster charge
      ($Q$).
\begin{linenomath}
\begin{eqnarray}
   {S_1}^2 &=& \frac{h^2 \cdot \tan^2\phi}{12} \;, \;\; {\rm and} \\
   N_{\rm eff}(1) &=& 1 \;\;\; {\rm by \;\; definition}.
\end{eqnarray}
\end{linenomath}
\item $N = 2$ \\
      Let the coordinates and charges of the clusters be $(x_1, Q_1)$ and
      $(x_2, Q_2)$. Their weighted-mean coordinate is given by
\begin{linenomath}
\begin{equation}
 X = \frac{x_1 Q_1 + x_2 Q_2}{Q_1 + Q_2} \;.
\end{equation}
\end{linenomath}
Its variance (${S_2}^2$) is given by
\begin{linenomath}
\begin{eqnarray}
 {S_2}^2 &\equiv& \left< (X - \left< X \right>)^2 \right> \nonumber\\
  &=& \left< \left( \frac{(x_1 - \left< X \right>)\cdot Q_1
        + (x_2 - \left< X \right>)\cdot Q_2}{Q_1 + Q_2} \right)^2 \right> \nonumber\\
  &=& \left<  \frac{(x_1 - \left< X \right>)^2\cdot {Q_1}^2
        + (x_2 - \left< X \right>)^2\cdot {Q_2}^2}{(Q_1 + Q_2)^2} \right> \nonumber\\
  &=& \left< (x - \left< x \right>)^2 \right> \cdot
           \left< \frac{{Q_1}^2 + {Q_2}^2}{(Q_1 + Q_2)^2} \right>  \nonumber\\
  &=& {S_1}^2 \cdot \left< \frac{(Q_1 + Q_2)^2 - 2Q_1Q_2}{(Q_1 + Q_2)^2} \right> \nonumber\\
  &=& {S_1}^2 \cdot \left( 1 - 2 \cdot 
                \left< \frac{Q_1Q_2}{(Q_1 + Q_2)^2} \right> \right) \nonumber\\
  &\geqq& {S_1}^2 \; / 2 \;.
\end{eqnarray}
\end{linenomath}
The third and fourth lines in the equation above are justified
since the variables $x$ and $Q$ are not correlated, whereas
the last line follows from
\begin{linenomath}
\begin{eqnarray} 
 \label{eqX}  Q_1Q_2 \; /(Q_1 + Q_2)^2 &\leqq& 1/4 \\
  \because \; (Q_1 - Q_2)^2 &=& (Q_1 + Q_2)^2 - 4Q_1Q_2 \geqq 0 \;. \nonumber 
\end{eqnarray}
\end{linenomath}
The equality in Eq. (\ref{eqX}) holds only when $Q_1 = Q_2$.
Therefore, in a general case addressed here 
\begin{linenomath}
\begin{equation}
 N_{\rm eff}(2) \equiv {S_1}^2 \;/{S_2}^2 < 2 \;.
\end{equation}
\end{linenomath}

\item $N = 4$ \\
\begin{linenomath}
\begin{eqnarray}
 X &=& \frac{x_1Q_1 + x_2Q_2 + x_3Q_3 +x_4Q_4}{Q_1+Q_2+Q_3+Q_4} \nonumber\\
   &=& \frac{x_1^\prime Q_1^\prime + x_2^\prime Q_2^\prime}
                    {Q_1^\prime + Q_2^\prime} 
\end{eqnarray}
\end{linenomath}
where
\begin{linenomath}
\begin{eqnarray}
x_1^\prime Q_1^\prime &\equiv& x_1Q_1 + x_2Q_2 \nonumber\\
x_2^\prime Q_2^\prime &\equiv& x_3Q_3 + x_4Q_4 \nonumber
\end{eqnarray}
\end{linenomath}
with $Q_1^\prime \equiv Q_1 + Q_2$ and  $Q_2^\prime \equiv Q_3 + Q_4$.
\begin{linenomath}
\begin{eqnarray}
 {S_4}^2 &\equiv& \left < (X - \left< X \right>)^2 \right> \nonumber\\
       &=& \left< (x^\prime - \left< x \right>)^2 \right> \cdot
          \left< \frac{{Q_1^\prime}^2 + {Q_2^\prime}^2}
                     {(Q_1^\prime + Q_2^\prime)^2} \right> \nonumber\\
  &=& \left< (x^\prime - \left< x \right>)^2 \right> \cdot
          \left( 1 - 2 \cdot \left< \frac{Q_1^\prime Q_2^\prime}
              {(Q_1^\prime + Q_2^\prime)^2} \right> \right) \nonumber\\
  &>& {S_2}^2 / 2 \;,\;\;{\rm with} \;\;
          {S_2}^2 \equiv \left< (x^\prime - \left< x \right>)^2 \right> \;. 
\end{eqnarray}
\end{linenomath}
Therefore
\begin{linenomath}
\begin{eqnarray}
 \frac{{S_2}^2}{{S_4}^2} &<& 2 \;,\;\;{\rm and} \\
 N_{\rm eff}(4) &<& 2N_{\rm eff}(2) \;.
\end{eqnarray}
\end{linenomath}
Consequently 
\begin{linenomath}
\begin{equation}
N_{\rm eff}(1) = 1,\; N_{\rm eff}(2) < 2,\;
            N_{\rm eff}(4) < 2 N_{\rm eff}(2), \;\; {\rm and \; so \; on.}
\end{equation}
\end{linenomath}
Thus $N_{\rm eff}(N) / N$ is expected to be a decreasing function of $N$.

\item $N = \infty$
\begin{linenomath}
\begin{eqnarray}
  X &=& \frac{\sum_{i=1}^N x_i Q_i}{\sum_{i=1}^N Q_i} \\
 {S_{\rm N}}^2 &\equiv& \left< (X - \left< X \right>)^2 \right> \nonumber\\
         &=&  \left< (x - \left< x \right>)^2 \right> \cdot  
 \left< \frac{\sum_{i=1}^N {Q_i}^2}{\left( \sum_{i=1}^N Q_i\right)^2} \right> \nonumber\\ 
   &\sim& {S_1}^2 \cdot \frac{\left< \sum_{i=1}^N {Q_i}^2 \right>}
                                          {N^2 \left< Q \right>^2} \nonumber\\
   &\sim& {S_1}^2 \cdot \frac{1}{N} \cdot
                  \frac{\left< Q^2 \right>}{\left< Q \right>^2} \nonumber\\
   &\sim& {S_1}^2 \cdot \frac{1}{N} \cdot \frac{\left< Q \right >^2
       + {\sigma_Q}^2}{\left< Q \right>^2} \nonumber\\
   &\sim& {S_1}^2 \cdot \frac{1}{N} \cdot (1 + F^\prime) 
\end{eqnarray}
\end{linenomath}
where the relative variance $F^\prime \equiv {\sigma_Q}^2\; / \left< Q \right>^2$ 
with $\sigma_Q$ being the standard deviation of the cluster charge,
including the fluctuations in cluster size, and in avalanche gain
for each electron in the cluster.
Actually 
\begin{linenomath}
\begin{equation}
F^\prime = F + \frac{1}{\left<n\right>} \cdot f 
\end{equation}
\end{linenomath}
with $F$ ($f$) being the relative variance of the cluster-size (avalanche-size)
fluctuation and $\left<n\right>$ the average cluster size.
Therefore
\begin{linenomath}
\begin{equation}
 \lim_{N \rightarrow \infty} \frac{N_{\rm eff}(N)}{N} = \frac{1}{1+F^\prime} 
  \sim  \frac{1}{1+F}
\end{equation}
\end{linenomath}
because $F$ ($\sim$ 2000 for argon \cite{ref3}) is much greater than 
$f$ ($\sim$ 1).
\end{enumerate}

\end{document}

%% file: authorlist.tex
%
%
\author[3]{M.~Kobayashi\corref{cor1}}
     \ead{makoto.kobayashi.exp@kek.jp}
     \cortext[cor1]{Corresponding author.
                           Tel.: +81 29 864 5379; fax: +81 29 864 2580.}
\author[7]{R.~Yonamine}
\author[2]{T.~Tomioka}
\author[1]{A.~Aoza}
\author[2]{H.~Bito}
\author[3]{K.~Fujii}
\author[1]{T.~Higashi}
\author[4]{K.~Hiramatsu}
\author[3]{K.~Ikematsu}
\author[1]{A.~Ishikawa}
\author[4]{Y.~Kato}
\author[1]{H.~Kuroiwa}
\author[3]{T.~Matsuda}
\author[2]{O.~Nitoh}
\author[2]{H.~Ohta}
\author[2]{K.~Sakai}
\author[5]{R.D.~Settles}
\author[1]{A.~Sugiyama}
\author[1]{H.~Tsuji}
\author[6]{T.~Watanabe}
\author[3]{H.~Yamaoka}
\author[4]{T.~Yazu}
%
%
\address[3]{High Energy Accelerator Research Organization (KEK), Tsukuba, 305-0801, Japan}
\address[7]{Graduate University for Advanced Studies, KEK, Tsukuba, 305-0801, Japan}
\address[2]{Tokyo University of Agriculture and Technology, Koganei, 184-8588, Japan}
\address[1]{Saga University, Saga, 840-8502, Japan}
\address[4]{Kinki University, Higashi-Osaka, 577-8502, Japan}
\address[5]{Max Planck Institute for Physics, DE-80805 Munich, Germany}
\address[6]{Kogakuin University, Hachioji, 192-0015, Japan}

%% file: Table1.tex
\begin{table}[htbp]
\begin{center}

\caption{\label{tb1:table1}
 Results of measurement and calculation}

\medskip

\begin{tabular}{|c||c|c||c|c|c|c|} \hline
 & \multicolumn{1}{|c|}{$\sigma_{\rm X0}$ ($\mu$m)} &
 \multicolumn{1}{|c||}{$n_{\rm eff}$} & 
 \multicolumn{2}{|c|}{$S_{\rm X00}$ ($\mu$m)} & 
 \multicolumn{2}{|c|}{$N_{\rm eff}$} \\
\cline{4-7}
$\phi$ ($^\circ$)& Measured & Measured & Measured & Calculated & Measured & Calculated \\
\hline \hline
\hspace{0.4mm} 0 & \hspace{2mm}96 $\pm$ \hspace{2.0mm}6 & 22.8 $\pm$ 0.8 & \hspace{2.4mm} $\equiv$ 0  & \hspace{1.7mm}   0 
& $-$ & 5.11 \\ \hline
\hspace{0.4mm} 5 & 124 $\pm$ \hspace{2.0mm}5 & 21.5 $\pm$ 0.7 & \hspace{2.4mm}  79.2 $\pm$ 10.2 & \hspace{2.7mm}  70.3
& 4.0 $\pm$ 1.1 & 5.12 \\ \hline
10 & 171 $\pm$ 14 & 24.5 $\pm$ 2.7 & \hspace{0.4mm} 142.1 $\pm$ 16.3 & \hspace{0.9mm} 141.2
& 5.1 $\pm$ 1.2 & 5.15 \\ 
\hline

\end{tabular} 

\end{center}
\end{table}

%% file: main.bbl
\newpage

\begin{thebibliography}{99}
\bibitem{ref1}  M.~Kobayashi, {\rm et al.},
                    Nuclear Instrumentation and Methods in Physics Research \\
                    A 641 (2011) 37.
\bibitem{refLC1}
	       The International Linear Collider,
               ILC Technical Design Report, available at \\
	$<$https://www.linearcollider.org/ILC/Publications/Technical-Design-Report$>$.
\bibitem{refLC2}
	       The Compact Linear Collider, Compact Linear Collider, 
               available at \\
               $<$http://clic-study.org/$>$.
\bibitem{ref2} Makoto~Kobayashi,
                    Nuclear Instrumentation and Methods in Physics Research \\
                    A 562 (2006) 136.
\bibitem{ref3} Makoto~Kobayashi,
                    Nuclear Instrumentation and Methods in Physics Research \\
	            A 729 (2013) 273.
\bibitem{ref4} R.~Yonamine, et al., 
	            Journal of Instrumentation 9 (2014) C03002.
\bibitem{ref5} H.~Fischle, J.~Heintze, B.~Schmidt,
                    Nuclear Instrumentation and Methods in Physics Research \\
	            A 301 (1991) 202.
\bibitem{ref6} A.~Sharma, F.~Sauli,
                    Nuclear Instrumentation and Methods in Physics Research \\
	            A 350 (1994) 470.
\bibitem{ref7} S.F.~Biagi,
                    Nuclear Instrumentation and Methods in Physics Research \\
	            A 421 (1999) 234.
\end{thebibliography}
